**Volatile Loss and Classification of Kuiper Belt Objects**

R.E. Johnson[1,2], A. Oza[3,4], L.A.Young[5], A.N.Volkov[6], C. Schmidt[1,3]

[1] Engineering Physics, University of Virginia, Charlottesville, VA 22904
[2] Physics, New York University, NY, NY 10003
[3] Department of Astronomy, University of Virginia, Charlottesville, VA 22904
[4] CNRS, LATMOS/IPSL, Université Pierre et Marie Curie, Paris, France
[5] SwRI, 1050 Walnut St., Boulder, CO 80302-5150
[6] Dept. of Mechanical Engineering, University of Alabama, Tuscaloosa, AL 35487

**Abstract**

Observations indicate that some of the largest Kuiper Belt Objects (KBOs) have retained volatiles in the gas phase, which implies the presence of an atmosphere that can affect their reflectance spectra and thermal balance. Volatile escape rates driven by solar heating of the surface were estimated by Schaller and Brown (2007) (SB) and Levi and Podolak (2009)(LP) using Jeans escape from the surface and a hydrodynamic model respectively. Based on recent molecular kinetic simulations these rates can be hugely in error (e.g., a factor of ~$10^{16}$ for the SB estimate for Pluto). In this paper we estimate the loss of primordial $N_2$ for several large KBOs guided by recent molecular kinetic simulations of escape due to solar heating of the surface and due to UV/EUV heating of the upper atmosphere. For the latter we extrapolate simulations of escape from Pluto (Erwin et al. 2013) using the energy limited escape model recently validated for the KBOs of interest by molecular kinetic simulations (Johnson et al. 2013). Unless the $N_2$ atmosphere is thin (<~$10^{18}N_2/cm^2$) and/or the radius small (<~200-300km), we find that escape is primarily driven by the UV/EUV radiation absorbed in the upper atmosphere rather than the solar heating of the surface. This affects the previous interpretations of the relationship between atmospheric loss and the observed surface properties. The long-term goal is to connect detailed atmospheric loss simulations with a model for volatile transport (e.g., Young, 2014) for individual KBOs.

**1. Introduction**

Pluto's atmosphere was discovered through stellar occultation about a quarter of a century ago (e.g., Elliot et al. 1989), and the Voyager spacecraft measured Triton's atmosphere (Yelle et al. 1995). Now a number of large icy objects, at or past the orbit of Neptune, appear to have surface volatiles suggesting that vapor-pressure supported atmospheres are likely present during all or part of their orbits. We call these the volatile-bearing Kuiper Belt Objects (KBOs), although one is a moon of Neptune, and three (Pluto, Eris, and Makemake) are dwarf planets. KBOs are known to exhibit a variety of surfaces. How much of this variety is due to different origins and processing, and how much is due to common processes is uncertain. As has been suggested often, the vapor-pressure supported atmosphere might provide a link. It controls the loss of the volatiles by escape, the appearance of radiation-induced chemical products on the surface, and the freshening of the surface through condensation. Here we correct the estimates of the loss of the primordial volatiles as described in Schaller and Brown (2007) (hereafter, SB) using Jeans escape from the surface and in Levi and Podolak (2009)(hereafter, LP) using an isentropic hydrodynamic model. This study is possible because of the growth in our knowledge of these bodies (e.g., Brown 2012) and recent molecular kinetic simulations of escape that re-examined both Jeans and hydrodynamic models for atmospheric escape (Volkov et al. 2011a,b; Tucker et al. 2012; Erwin et al. 2013; Johnson et al. 2013).

In this work we focus on loss of $N_2$, the most volatile species of the three ($N_2$, $CH_4$, CO) known to be present on Pluto and Triton and examined in SB and LP. Of course, these volatiles cannot be treated separately as a small amount of CO can be an important cooling agent and even a few percent of $CH_4$ dominates the heating of the upper atmosphere (Krasnopolsky and Cruikshank 1999). Therefore, we estimate the loss of the initial inventory of $N_2$ from a number of KBOs having various sizes and orbital parameters assuming that a trace amount of $CH_4$ is present when the atmosphere is predominantly $N_2$. We first review the data on the KBOs of interest, then describe and critique the previous estimates for atmospheric escape from KBOs and present improved approximate description based on recent molecular kinetic and fluid simulations. Finally, new estimates for the retention of $N_2$ are presented and discussed in light of the available observations.

## 2. Properties of Large KBOs

We consider the eight largest bodies in the Kuiper Belt, including Neptune's moon, Triton, and Pluto's moon Charon. We exclude Haumea, since it is likely that the volatile retention is influenced more by collisions than by escape (Brown et al. 2007a). Table 1 lists the relevant properties: the radius of the body ($r_0$), its bulk density ($\rho$), and Bond albedo (*A*, more relevant than the geometric albedo for thermal balance); observations of their surfaces and atmospheres; and the semimajor axis (*a*) and eccentricity (*e*) of their orbits. For a number of these bodies, even Pluto, the densities and radii are not known accurately. For three KBOs we list a range of densities, and for Pluto, for which the mass is well constrained but not the radius, we list the parameters used in our recent simulations for consistency.

It is seen that these KBOs fall into classes in terms of size, albedo, heliocentric distances, and volatiles detected. Using new estimates of the atmospheric loss rate we determine the net loss of $N_2$ from the larger, brighter bodies (Triton, Pluto, Eris, Makemake), which have retained a large fraction of their initial volatile inventory. We also re-evaluate the medium-sized bodies, 2007 OR10 and Quaoar, which have been suggested to have lost most or all of their initial $N_2$ component (SB), with their low albedos, a result of radiation processing of gaseous or condensed methane (e.g., Johnson 1989; Strazzulla et al. 2003). Charon and Sedna are outliers with intermediate albedos, the former having rapidly lost its primordial volatiles but at present acquires nitrogen from Pluto (Tucker et al. 2014), and the latter in a very distant orbit with a very thin atmosphere having retained its volatiles. The data in Table 1 is used below to calculate atmospheric loss rates for KBOs in their current orbits. The role of the evolution of the orbits and albedos will be considered in the future.

## Table 1. KBO Properties Ordered by Radius

| KBO[a] | $a$ (AU) | $E$ | $r_0$ (km) | $\rho$ (g cm$^{-3}$) | $A$ | Surface Composition | Atmosphere | References |
|---|---|---|---|---|---|---|---|---|
| Triton | 30.04 | 0.01 | 1150 | 2.06 | 0.80[b] | $N_2$, CO, $CH_4$, $H_2O$, $CO_2$, $C_2H_6$; Heterogeneous. | ~14 μbar, global @ 30 AU. $N_2$, trace $CH_4$, CO. Albedo suggests atmospheric freshening | Thomas 2000; McKinnon 1995; Stansberry et al. 1990; Grundy et al. 2010; DeMeo et al. 2010; Lellouch et al. 2010 |
| Pluto[c] | 39.17 | 0.24 | 1153 | 2.05 | 0.67 | $N_2$, CO, $CH_4$, tholins, $C_2H_6$; little/no $H_2O$; Heterogeneous. | ~17 μbar, global @ 30-32 AU. $N_2$ trace $CH_4$, CO. Albedo suggests atmospheric freshening | Lellouch et al. 2009, 2011; Buie et al. 2006; Grundy et al. 2014; Holler et al. 2014 |
| Eris | 67.94 | 0.43 | 1163 | 2.52 | 0.55 | $N_2$, $CH_4$; | <1 nbar @ 96 AU. Albedo suggests atmospheric freshening. | Sicardy et al. 2011; Tegler el al. 2012; Abernathy et al. 2009 |
| Makemake | 45.57 | 0.16 | 733 | 1.6-2.1[d] | 0.62 | $N_2$, $CH_4$, $C_2H_6$; Heterogeneous. | <12 nbar @ 52 AU. Albedo suggests atmospheric freshening. | Ortiz el al. 2012; Lim et al. 2010; Stansberry et al. 2008; Tegler et al. 2008; Brown et al. 2007b |
| 2007 OR$_{10}$ | 66.88 | 0.50 | 640 | 1.6-2.1[d] | 0.06 | $CH_4$ (inferred); $H_2O$ | No constraints @85-87 AU. | Santos-Sanz et al. 2012; Brown et al. 2011 |
| Charon | 39.17 | 0.24 | 606 | 1.72 | 0.25 | $H_2O$, $NH_3$ | <110 nbar $N_2$ and <15 nbar $CH_4$ @ 31 AU. | Person et al. 2006; Brozović et al. 2014; Lellouch et al. 2011; Cook et al. 2007; Gulbis et al. 2006 |
| Quaoar | 43.18 | 0.04 | 534 | 2.18 | 0.07 | $N_2$ (v. tentative), $CH_4$, $H_2O$, $C_2H_6$, "dark material" | <20 nbar $CH_4$ @ 43 AU. | Fornasier et al. 2013; Dalle Ore 2009; Schaller & Brown 2007b; Braga-Ribas et al. 2013; |
| Sedna | 541.79 | 0.86 | 498 | 1.6-2.1[d] | 0.19 | $H_2O$, serpentine, $N_2$ (trace), $CH_4$ (trace), $C_2H_6$ (trace), tholins. | Occultation, with no atmospheric analysis @ 87 AU. | Pal et al. 2012. Emery et al. 2007; Barucci et al. 2010; Braga-Ribas 2013; |

[a] a, semimajor axis; $e$, orbital eccentricity; $r_0$, average surface radius; $\rho$, mass density; A, bond albedo
[b] south polar cap.
[c] $N_2$-rich terrain; values of $r_0$ and $\rho$ used in Erwin et al. 2013.
[d] No known satellite, density assumed Charon-like to Pluto/Triton-like.

## 3. Atmospheric Escape Models

We first discuss several approximations to the escape of volatiles from KBOs, including Jeans escape (used in Schaller & Brown 2007; SB), Parker's hydrodynamic model of isentropic outflow (used in Levi and Podolak, 2009; LP), and energy-limited escape due to heating of the upper atmosphere (Johnson et al. 2013). Their applicability depends on how tightly bound and rarefied the atmosphere is. These qualities are indicated by the surface values of the Jeans parameter, $\lambda_0$, and on the atmospheric column density, $N_0$. In addition to directly affecting the escape rate, the column density determines the opacity of the atmosphere to UV insolation and, therefore, determines the atmospheric heating. In this work, we consider a nitrogen-dominated atmosphere, the most volatile of the three species considered in SB and LP. Following SB we assume an average surface temperature, $T_0$, in radiative equilibrium: $\varepsilon \sigma T_0^4 = [F_{IAU}(1-A)/R^2]/4$, where $\varepsilon$ is the emissivity, $F_{IAU}$ is the solar insolation at 1 AU, $A$ is the bolometric Bond albedo, $R$ is the orbital position in AU, $\sigma$ the Stefan-Boltzmann constant and the factor of four accounts for global averaging. The surface temperature is, of course, influenced by thermal inertia and latent heat of sublimation (Young and McKinnon 2013), which will be addressed in later work.

The Jeans parameter is the ratio of potential energy, $U = GM_{KBO}\,m/r$, to thermal energy, $kT$, which at the surface, $r = r_0$, is written as $\lambda_0 = U_0/(kT_0)$. Here $U(r)$ is the gravitational binding energy for a molecule of mass $m$ to a KBO of mass $M_{KBO}$, $G$ is the gravitational constant, and $k$ is the Boltzmann constant. The Jeans parameter is also related to the scale height at the surface, $H_0$, by $H_0 = r_0/\lambda_0$, with a large $\lambda_0$ indicating a tightly bound atmosphere. Writing a column density as $N_0 = n_0 H_0$, then $N_0 = P_{vap}(T_0)/(mg_0)$ using the equation of state, $P_0 = n_0 kT_0$, and hydrostatic equilibrium, $P_0 = P_{vap}(T_0)$, where $g_0 = U_0/r_0$ is the surface gravity and $P_{vap}$ is the vapor pressure of the $N_2$ frost on the surface. Moderately bound escaping atmospheres have correction terms of order $1/\lambda_0$ (Elliott and Young, 1992) which we ignore. Although $N_0$ underestimates the zenith column, we use it in Fig. 1 as a proxy for the surface pressure to characterize the KBO atmospheres. It is seen that the KBOs of interest have moderately bound atmospheres, with $\lambda_0$ in the range of 10 – 100. Because the surface pressure depends very strongly on the temperature of the $N_2$ ice, $N_0$ can vary dramatically over a KBO's season.

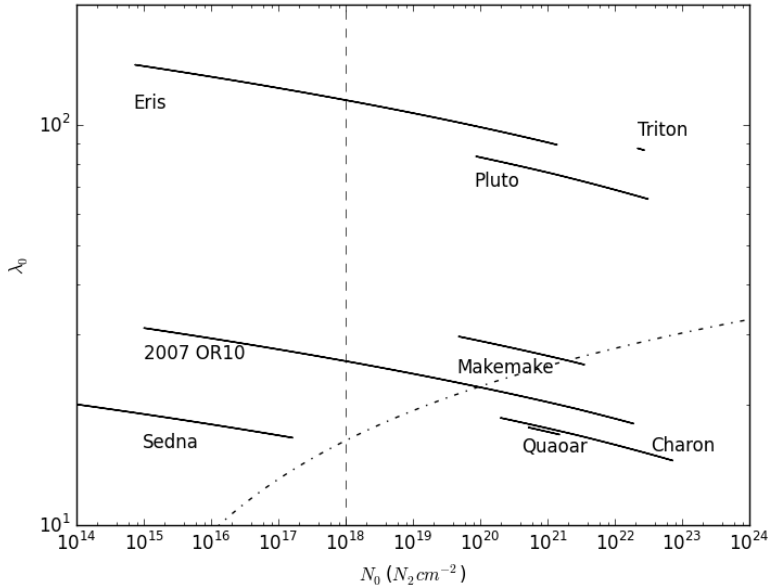

Fig. 1 Jeans parameter at the surface, $\lambda_0$, vs. column density, $N_0$, for an $N_2$ atmosphere on a number of KBOs using the data in Table 1 for $a$, $e$, $r_0$, and $\rho$ but with $A = 0.67$ when the KBO still has a significant $N_2$ atmosphere: $\lambda_0 = U(r_0)/kT_0$ with $N_0 = P_{vap}(T_0)/(mg_0)$ determined from the $N_2$ vapor pressure at a particular point in a KBO's orbit and $T_0$, a surface temperature determined from $\varepsilon \sigma T_0^4 = [F_{IAU}(1-A)/R^2]/4$ with emissivity, $\varepsilon = 1$, and $R$ the distance from the Sun. To the right of the dashed line, $N_0 > N_c = 10^{18}$ $N_2/cm^2$, so

that the CH$_4$ component is sufficient to absorb the UV as discussed in section 3.2. The dot-dashed: (simulated loss rate/ SJ) = 1 in Fig. 2; therefore, Eq. 1b *overestimates* the surface-heating-induced escape rate below this curve. For Makemake, 2007OR10, and Sedna with uncertain densities in Table 1 we used $\rho$ = 1.8g/cm$^3$.

Accurately modeling escape from an atmosphere of a KBO heated at the surface by the solar radiation and with the upper atmosphere heated in the UV/EUV requires detailed knowledge of the composition, chemistry and volatile transport (e.g., Zhu et al. 2014; Young et al. 2012). Therefore, consistent with our present understanding of the KBOs, simpler models are still warranted. We first critique the models for escape driven by surface heating used in SB and LP based on recent molecular kinetic simulations (Volkov et al. 2011a,b; Volkov and Johnson 2013; Volkov et al. 2013; Johnson et al. 2013a,b). These simulations correctly treat the transition in an atmosphere from the collision-dominated regime into the collisionless regime where escape occurs. This is followed by a description of a model that accounts for the short wavelength heating of the upper atmosphere. We then use a simple estimate for the combined effect in order to recalculate the loss rates.

### *3.1 Escape Driven by Surface Heating*

Atmospheric loss from KBOs driven *only* by the solar heating of the surface has been estimated in both LP and SB. The former used Parker's isentropic fluid model of escape from KBOs driven by the surface temperature. They found solutions that went sonic for a number of KBOs. However, in an N$_2$ atmosphere driven *only* by surface heating, molecular kinetic simulations have shown that sonic solutions occur for $\lambda_0$ much smaller than those in Fig. 1 (e.g., Volkov et al 2011a, Fig. 3; Volkov and Johnson 2013). Therefore, in the remainder we compare to the work in SB.

Atmospheric loss was estimated in SB by evaluating the Jeans escape rate at the surface rather than at the exobase. In this way they avoided determining the exobase properties and they could treat the loss rates for three species (N$_2$, CH$_4$, CO) separately. For a body of radius, $r_0$, having a uniform surface temperature, $T_0$, the surface *sublimation rate* for molecules with mass $m$ and vapor pressure $P_{vap}$ is given by

$$(dM/dt)_0 = 4\pi r_0^2 m[P_{vap}(T_0) / (2\pi m kT_0)^{0.5}] \qquad (1a)$$

multiplied by the fractional coverage of the surface by the volatile. For an atmosphere for which escape can occur directly from the surface, which we will call the surface-Jeans (SJ) estimate, one writes

$$(dM/dt)_{SJ} = (dM/dt)_0 (1+\lambda_0) \exp(-\lambda_0) \qquad (1b)$$

SB suggested that if heating of the upper atmosphere is ignored, this rate is a lower bound to the actual escape rate. For a hydrostatic, isothermal atmosphere with no UV absorption in the upper atmosphere, this is the case as the ratio of the Jeans rate at the exobase to the SJ rate is $\sim r_x/r_0$, where $r_x$ is the exobase radius. Of course, atmospheres experiencing escape on such bodies are *not* isothermal. Therefore, the estimate in Eq. 1b was tested by molecular kinetic simulations of atmospheric escape driven only by the surface heating (Volkov et al. 2011a,b) for a range of $\lambda_0$ and surface values of the radial Knudsen number, $Kn_0$, which is a measure of the atmospheric thickness. It is the ratio of the mean free path between collisions, $l_0 = 1/n_0\sigma_{eff}$, to a length scale, here $r_0$: $Kn_0 = 1/n_0\sigma_{eff}r_0$, where $\sigma_{eff}$ is an effective cross section for collisions between atmospheric molecules. Using definitions above we can also write $Kn_0 = 1/[\lambda_0\sigma_{eff}N_0]$. We primarily use the column density, $N_0$, as our parameter in discussing the KBO atmospheres.

In Fig. 2 we give the ratio of the escape rate of $N_2$ from a KBO atmosphere to the SJ rate in Eq. 1b using two models. Because molecular kinetic simulations are computationally intense for thick atmospheres and large $\lambda_0$, ratios are given in Fig. 2b only for $\lambda_0 = 10$ and 15 at the smaller $N_0$ of interest in Fig. 1. These ratios increase as ~ $1/Kn_0^a$ at small $N_0$ for each $\lambda_0$ (Volkov et al. 2011b). At large $N_0$ and $\lambda_0$, the fluid-Jeans (FJ) approximation has been shown to be a reasonable approximation, which underestimates the molecular kinetic simulation by a factor < ~2 at the largest $N_0$ and $\lambda_0$ (e.g., Volkov et al. 2011a,b). In the FJ model the one-dimensional equations for an inviscid, heat conducting gas are solved iteratively applying the Jeans mass and energy escape rates at the upper boundary and taking into account the non-zero gas flow speed (Yelle 2004; Volkov et al. 2011a,b). In Fig. 2a we give the ratio of the FJ rate to the SJ rate in Eq. 1b for two different locations of upper boundary to account for collisions in the exosphere (e.g., Tucker and Johnson 2009). Results are shown for $l_e / H_e = 1$ and 3 where $l_e$ and $H_e$ are the mean free path and scale height at the upper boundary from which escape is presumed to occur. The choice of this boundary affects the ratio at small $N_0$ but not at the larger values. It is seen in Fig. 2b, that the FJ rate roughly approaches the molecular kinetic model at the larger $N_0$ (Volkov et al. 20011a). This is the conduction dominated regime, the so-called Fourier regime in Gruzinov (2011), in which the ratio in Fig 2a at large $N_0$ varies as $Kn_0$ for fixed $\lambda_0$: ~ $b[Kn_0 \exp(\lambda_0) / \lambda_0^c]$. It is also seen in Fig. 2b that the FJ rate underestimates the molecular kinetic escape rate at small $N_0$, although increasing the height of the upper boundary reduces that difference somewhat. In the absence of additional kinetic simulations the thin dashed lines, described in the caption, are used to estimate the surface heating contribution to escape.

The ratios in Fig. 2a and b indicate that the SJ rate *is* a lower bound to the escape rate at small $N_0$ for each $\lambda_0$, but significantly *overestimates* the escape rate at large $N_0$. For instance, for a KBO with $\lambda_0 \sim 10$, the SJ estimate is a lower bound to the thermally driven rate for surface column densities, $N_0 < \sim 10^{16}\,N_2/cm^2$. The change from the SJ rate being a lower bound to being an upper bound moves to larger $N_0$ as $\lambda_0$ increases, also indicated by the dot-dash line in Fig.1. Therefore, for Quaoar and Charon, as well as Makemake and 2007 OR10 near perihelion, the SJ rates significantly overestimates escape driven by surface heating. On the other hand simulations for Pluto show that the SJ *underestimates* the escape rate by many orders of magnitude (~$10^{16}$ near perihelion with $\lambda_0 \sim 60$: in Fig. 1). That is, rather than losing ~1.6 x $10^{11}$ $N_2$/s as estimated using Eq. 1b, Pluto loses ~ 2.6 - 3.5x$10^{27}$ $N_2$/s at solar medium conditions (Erwin et al. 2013). Therefore, the SJ rate is neither a good approximation nor a lower bound except over a narrow range of $N_0$ and $\lambda_0$. The large loss rate calculated for Pluto is due to the UV/EUV radiation absorbed in the upper atmosphere (e.g., Erwin et al. 2013) as described below. Therefore, reasonably accurate estimates must include both heating of the surface and direct heating of the upper atmosphere by the short wavelength radiation. Below we describe an approximation for the UV/EUV driven escape and combine that model with the results in Fig. 2.

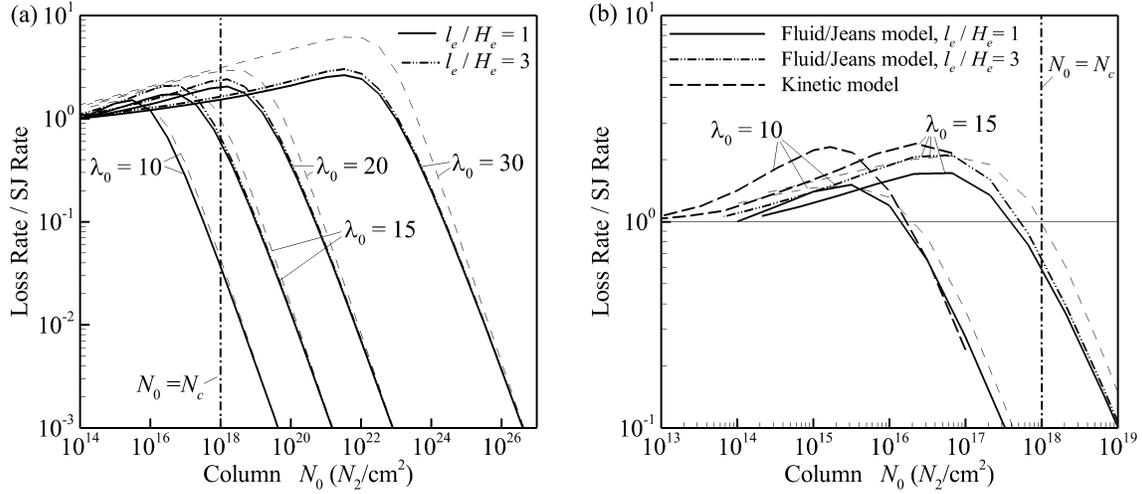

Fig.2 Calculated loss rate from an $N_2$ atmosphere produced by surface sublimation scaled to the SJ rate in Eq. 1b for a number of $\lambda_0$ as a function of $N_0$; (vertical dash-dot line) $N_0 = N_c$; (a) FJ model: hydrodynamic simulation iteratively applying Jeans conditions at an upper boundary from which escape occurs determined by the mean free path between $N_2$ collisions, $l_e$, and the local scale height, $H_e$: (solid lines) upper boundary at $l_e / H_e = 1$, (dash-two dots) $l_e / H_e = 3$; (b) FJ model as in (a) compared to molecular kinetic simulations (dashed lines) which roughly merge with the FJ result as $N_0$ increases. Expansion of the work in Volkov et al 2011b and 2013; instead of $N_0$ they used: $Kn_0 = 1/n_0\sigma_{eff}(T_0)r_0 \sim 1/[\lambda_0\sigma_{eff}(T_0)N_0]$, where $\sigma_{eff}(T_0) = 2^{1/2}\sigma(T_0)$ with $\sigma(T_0)$ the collision cross section of pseudo-Maxwellian molecules at temperature $T_0$: although slowly dependent on $T_0$ we use an average value to determine $N_0$: $\sigma_{eff} = 1.0 \times 10^{-14}$ cm$^2$ for $N_2 + N_2$ collisions. Writing the ratios at small $N_0$ as $R_1 \sim 1/Kn_0^a$ and at large $N_0$ as $R_2 \sim b[Kn_0 \exp(\lambda_0)/\lambda_0^c]$ the simulated ratios are approximated as $\sim [R_1^{-1} + R_2^{-1}]^{-1}$: solid lines for FJ in Fig. 2a are well fit (<~20%) using $a = 0.45$, $b = 70$ and $c = 2.55$; (thin dashed lines) to roughly account for the difference between the kinetic and FJ in Fig. 2b, $a = 0.09$ is used in $R_1$; in the absence of kinetic results at larger $N_0$ and $\lambda_0$ the expression giving thin dashed lines is used in the text to estimate the surface heating contribution to escape.

### 3.2 UV/EUV Driven Escape

When the radiation absorbed in the upper atmosphere is important, unlike in SB and LP, the volatiles cannot be treated separately. That is, UV absorption by the small concentration of $CH_4$ in Pluto's atmosphere dominates the upper atmospheric heating. This largely results from $CH_4$ photolysis by Lyman-α, a wavelength at which pure nitrogen atmospheres are transparent. If a KBO has retained a significant column of $N_2$, then the less volatile $CH_4$ will be present as a trace species, as is the case at Triton and Pluto. Therefore, for KBOs having various sizes and orbital parameters we estimate the loss of primordial nitrogen from an $N_2$ atmosphere with a small fraction of $CH_4$, which would be typical of early KBO atmospheres.

In simulations of escape from Pluto, a detailed description of the atmosphere as a function of altitude was constructed in response to the solar heating at $R \sim 30$AU. It was also shown that the so-called energy limited (EL) loss rate gave a reasonable estimate of the accurate globally averaged escape rate (Erwin et al. 2013). The EL approximation to the escape rate, applied in early studies of Pluto's atmosphere (Watson et al. 1982), and more recently to exoplanet atmospheres (Lammer et al. 2009), was tested for the first time using molecular kinetic simulations (Johnson et al. 2013). In this estimate it is assumed that the energy absorbed in the tenuous regions of the upper atmospheres can be removed more efficiently by escape than by downward thermal condition or radiation to space (e.g., Erwin et al. 2013; Fig. 4). In its simplest form, the globally integrated escape rate is approximated as

$$(dM/dt)_{EL} \sim c' \, m \, Q/U(r) \tag{2a}$$

Here $Q$ is the net atmospheric heating rate, $U(r)$ is the average gravitational energy of the molecules evaluated at $r$, a radius below the peak in the heating profile, and $c'$ is a factor that is the order of unity. Using $c' \sim 1$, the expression in Eq. 2a was shown to reasonably approximate the results of molecular kinetic simulations for a range of upper atmospheric heating rates, $Q$ (Johnson et al. 2013; Erwin et al. 2013; Tucker et al. 2012). It was also shown that this approximation is most accurate when the escape rate is large but Jeans-like rather when the heated upper atmosphere goes sonic, opposite to what was assumed for years (Johnson et al. 2013). For very high heating rates in which the atmosphere has a sonic point below the exobase it overestimates the loss rate, which is also the case when $Q$ becomes small.

The EL mass loss rate, with $c' = 1$ in Eq. 2a, can be written as

$$(dM/dt)_{EL} \sim mQ/U(r) = \eta <F_{EUV/UV}> 4\pi r_a^2 /(GM_{KBO}/r) \tag{2b}$$

Here $<F_{EUV/UV}>$ is the *globally averaged* energy flux absorbed in the UV and EUV, $\eta$ is the heating efficiency, and $r_a$ is an effective radius associated with the peak in the UV absorption. Typically, $r_a$ is assumed to occur near or below the mean absorption depth for the EUV/UV, which is well below the exobase, $r_x$, and should be well above the physical surface, $r_0$. The energy in the denominator occurs because molecules lost to space must overcome gravity and, in steady state, be replaced from a radius $r$ below $r_a$.

Since heating expands the upper atmosphere, both $r_a$ and $r_x$ increase relative to $r_0$ as $Q$ is increased, which enhances the KBO's cross sectional area for absorption of radiation. The peak in the UV heating at Pluto occurred at $\sim 1.2$-$1.3 r_P$, depending slowly on the solar conditions where $r_P$ is Pluto's radius. Using these values to estimate $r$ and $r_a$ in Eq. 2b gave a reasonable approximation for both solar minimum and maximum conditions (Erwin et al. 2013: Table 4). On the other hand, using $r = r_P$ in Eq. 2b, the escape rates were *underestimated* by $\sim 1.3$ and $\sim 1.5$ at solar minimum and maximum respectively. Therefore, in scaling the UV/EUV heating and the escape from KBOs to the rates in Erwin et al. (2013), as described below, we are assuming $r \sim r_a \sim 1.2$-$1.3 r_0$ in Eqs. 2a & b. Since $M_{KBO} = (4\pi r_0^3/3)\rho_{KBO}$, where $\rho_{KBO}$ is the mass density, the result in Eq. 2b is seen to depend on the density and not on $r_0$ as noted often.

Although more detailed expressions for the EL rate have been discussed, in which corrections for thermal conduction and heat capacity are included (e.g., Lammer et al. 2009), we have shown that the accuracy depends primarily on the accuracy of the heating rate, $Q$, as estimated in Eq. 2b the effective cross section for absorption in the EUV by $N_2$ $\sim 0.91 \times 10^{-17} cm^2$ and that for absorption in the UV by $CH_4$ is $\sim 1.8 \times 10^{-17} cm^2$ (Krasnopolsky and Cruikshank 1999). To achieve an optical depth of $\sim 1$ in the UV requires a line of sight column of $\sim 5.5 \times 10^{16}$ $CH_4/cm^2$. Although the mixing ratios are uncertain, by scaling the heating rate to that used in the simulations of Pluto's upper atmosphere, in which a $CH_4$ mixing ratio of ~2-3% was assumed, requires an atmospheric column of ~ $2 \times 10^{18}$ $N_2/cm^2$ to achieve unit optical depth in the UV. The globally averaged heating rate is typically estimated using an average solar illumination angle of ~ $60°$ (e.g., Strobel et al. 1996) in which case a radial column density > ~ $1 \times 10^{18}$ $N_2/cm^2$ is very roughly required to obtain an average optical depth of unity in the UV. This is more than sufficient to achieve optical depth unity for $N_2$ absorption in the EUV. Therefore, in applying the EL estimate of the UV/EUV-induced $N_2$ loss rate, we require the radial column of $N_2$ at the surface must be greater than a critical column density, $N_c$ ~ $10^{18}$ $N_2/cm^2$ for absorption of the UV and to limit cooling by conduction to the surface. Because our parameter, $N_0$, underestimates the radial column by ~ 10 to 20% in going from $\lambda_0$ ~ 10 ~ 30 we use $N_0 > N_c$ as a conservative criterion for applying our calculation of the upper atmospheric heating effect described below.

Since the EL expression in Eq. 2b is linear in $Q$, we calculate the loss of $N_2$ by scaling to the recent simulations for Pluto. Therefore, as in Erwin et al. (2013) we ignore the loss of $CH_4$, but require that the column of $CH_4$ is sufficient to absorb the UV. In this way we are scaling the heating rates to those at 30AU in Krasnopolsky and Cruikshank (1999) and Krasnopolsky (1999). They give an effective, globally averaged $CH_4$ heating rate of $1.4 \times 10^{-3}$ ergs/cm$^2$/s and $N_2$ heating rate of $0.29 \times 10^{-3}$ ergs/cm$^2$/s at 30AU for solar medium conditions. The estimate for $CH_4$ uses a heating efficiency, $\eta$ ~ 0.5, and includes the Lyman-$\alpha$ resonantly scattered by the interplanetary medium, which predominantly illuminates the solar facing hemisphere. Although the heating efficiency, in Eq. 2b, is sensitive to the composition, they find similar efficiencies for pure $CH_4$ and for $CH_4$ in $N_2$. The $N_2$ rate is due to direct sunlight shortward of 800Å with $\eta$ ~ 0.25. The EUV heating rate for absorption by $N_2$ is direct from the sun and relatively unattenuated. Therefore, $Q$ scales as $R^{-2}$ for a pure $N_2$ atmosphere where $R$ is the radial distance from the sun. Due to Lyman-$\alpha$ absorption and scattering, the radial dependence of the UV heating of $CH_4$ is more complicated. It consists of the direct solar flux irradiating the disk, which is gradually attenuated in the outer solar system, the scattered interplanetary Lyman-$\alpha$, which irradiates KBOs from all directions, but with a larger flux from the direction of sun and a slower decay in $R$ (Gladstone 1998; Quémerais et al. 2013). In addition, there is an integrated stellar source that predominantly illuminates from south of the ecliptic. Because we are considering globally averaged loss rates and average solar conditions, we approximate the sum of the direct and scattered solar UV by scaling to the heating rate used for Pluto at 30AU for solar medium conditions, $Q_p$. To this we add a constant stellar Lyman-alpha background which eventually dominates at large $R$, and write the net heating as $Q \sim Q_p [(30/R)^2 + 0.09)]$. Therefore, in scaling to the simulations for the $N_2$ mass loss rate from Pluto at 30AU, $(dM/dt)_P$, we use it in the following

$$(dM/dt)_{EL} \sim (\rho_P / \rho_{KBO}) [(30/R)^2 + 0.09] (dM/dt)_P \qquad (3)$$

Here $\rho_P = 2.04$ g/cm$^3$ and $(dM/dt)_P = 120$ kg/s are the density and mass loss rate from Erwin et al. (2013), $\rho_{KBO}$ is the density of the KBO and $R$ is the radial distance in AU. Below we combine the estimates for escape driven by heating of the surface and radiation absorption in the atmosphere.

*3.3 Combined Surface and Upper Atmosphere Heating*

Loss from KBO atmospheres can in principal be driven by solar heating of the surface or a deep haze layer and by the UV/EUV heating of the upper atmosphere as discussed above. Zhu et al. (2014) combined the upper atmosphere model of Erwin et al. (2013) with an approximate lower atmosphere model. Because they found a slightly larger (~30%) loss rate and escape due only to the surface temperature is negligible, synergy between surface heating of the lower atmosphere and the direct heating of the upper atmosphere might contribute. Since that result is now being re-examined, the escape rates in Erwin et al. (2013) will be used here. We also note that the EL approximation to escape driven by upper atmospheric heating can overestimate the escape rate if the outward gas flow goes sonic or $Q$ becomes small. Based on the criterion in Johnson et al. (2013), sonic escape due to atmospheric heating does not occur for the KBOs of interest and, as $Q$ becomes small, the surface heating dominates. Therefore, at each orbital position, $R$, we use the larger of loss due to surface heating, *(dM/dt)$_S$*, and upper atmospheric heating, *(dM/dt)$_U$*, as a conservative estimate of the net escape rate from a KBO. For the latter we use the EL approximation in Eq. 3. For *(dM/dt)$_S$* we improve on the SJ rate in Eq. 1b using the very rough analytic estimate in Fig. 2a and b as given in the caption. The surface heating rate is very sensitive to the albedo, which might change over time, and both *(dM/dt)$_S$* and *(dM/dt)$_U$* depend on the mass density, which is not well constrained for a number of the KBOs in Table 1. The estimate of *(dM/dt)$_U$* is subject to the two constraints: the surface sublimation rate in Eq. 1a must exceed the escape rate, which appears to always be the case for the KBOs examined, and there must be a sufficient column of gas ($N_0 > N_c$) to account for the $CH_4$ contribution to the heating as indicated in Fig.1. To roughly indicate the uncertainties in this estimate, we also use an ad hoc model in which the loss processes are added in the form, *(dM/dt) ~ (dM/dt)$_S$ + f$_{UV}$ (dM/dt)$_U$*, using $f_{UV} \sim [1 - \exp(-\alpha N_0/N_c)]$. In this way $f_{UV}$ acts to gradually cut-off the UV/EUV contribution as the atmosphere becomes thin: i.e., as $N_0 \rightarrow 0$, $f_{UV} \rightarrow (-\alpha N_0/N_c)$ reducing the heating rate. After presenting the results below we conclude by discussing their relevance.

**4. Results**

We first find the range of radial distances, $R$, and radial size, $r_0$, for which the estimated escape rate driven by UV/EUV heating dominates the estimate of the escape rate driven by surface heating: *(dM/dt)$_U$ >(dM/dt)$_S$* indicated by the shaded areas in Fig. 3. We used the same average mass density, $\rho_{KBO}$ = 1.8g/cm$^3$, as that used in SB, and $T_0$ is determined for two Bond albedos, $A$, and unit emissivity. The nearly vertical solid and dashed lines (A = 0.67 & 0.1 respectively) in Fig. 3 indicate where the EL estimate of *(dM/dt)$_U$* from Eq. 3 equals the *(dM/dt)$_S$* estimate. The latter is obtained by multiplying the SJ rate in Eq. 1b by the rough analytic approximation to ratios in Fig. 2. The nearly horizontal lines indicate when $N_0 = N_c$ for each albedo. Since the sublimation source rate is adequate for resupplying the atmospheres of the KBOs of interest, it is not indicated. The radial size and orbital range for the KBOs in Table 1 are indicated by labeled vertical lines. The shaded light-grey area is the region for which *(dM/dt)$_U$ >(dM/dt)$_S$* assuming $A$ = 0.67 for a body with a significant $N_2$ atmosphere. The extension of that region for a much darker surface ($A$ = 0.1) is indicated by the dark grey region.

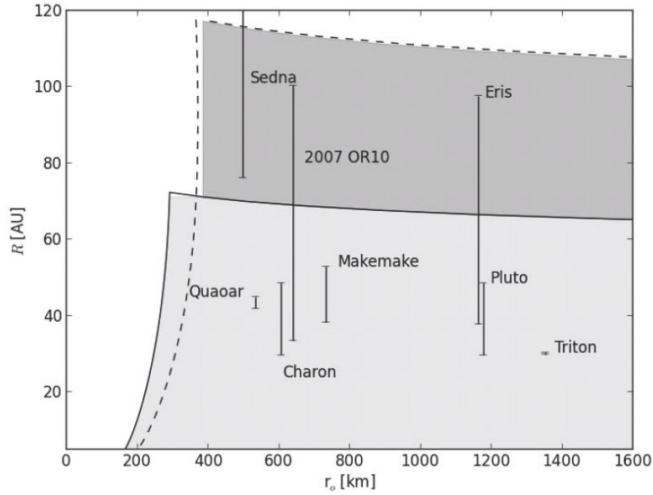

Fig. 3 Distance from sun, $R$ in AU, vs. radius, $r_0$ in km, for KBOs having an early $N_2$ atmosphere with a small fraction of $CH_4$ using an average density: $\rho_{KBO} = 1.8 g/cm^3$ as in SB. Solid lines ($A = 0.67$): (vertical) the estimated value of $r_0$ at each $R$ for which $(dM/dt)_S$ equals $(dM/dt)_U$; (horizontal) $N_c \sim 10^{18} N_2/cm^2$; $(dM/dt)_U$ dominates in light grey region. Dashed lines, same for ($A = 0.1$); dark grey indicates the extension of the region in which the $(dM/dt)_U$ estimate dominates for $A = 0.1$. The vertical lines indicate the orbital range of each the KBO in Table 1.

Even allowing for the fact that the analytic form in Fig. 2 might overestimate the surface contribution for relevant $N_0$ and $\lambda_0$, it is seen in Fig. 3 that except for Sedna, the evolution of the nitrogen component of the atmospheres of the KBOs of interest is either dominated by or to a large extent determined by escape driven by upper atmospheric heating. This is the case unless the body is small, $r_0 < \sim 300$km, or the atmospheric column becomes too thin to fully absorb the UV ($N_0 < N_c$). If the surface has a relatively bright frost ($A = 0.67$: solid lines) the atmosphere is thin on Sedna over its full orbit, but only beyond ~70AU for Eris and 2007OR10. In their very eccentric orbits, the dominant mass loss process changes with their distance from the sun during the time in which they retain significant $N_2$ to form a bright surface frost. In Fig. 4 we show this explicitly for 2007 OR10 again using an average density, $\rho_{KBO} = 1.8 g/cm^3$ and the two different albedos to calculate $(dM/dt)_S$. The solid and dot-dash lines are for $A = 0.67$ and 0.1 respectively using the simple cut-off between $(dM/dt)_S$ and $(dM/dt)_U$. The dashed lines in Fig. 4 is the mass loss rate calculated using $A = 0.67$ and $f_{UV}$, the gradual cut-off with $\alpha = 0.1$. Consistent with Fig. 3, $(dM/dt)_U$ dominates not only at the smaller values of $R$ but also dominates the net loss rate.

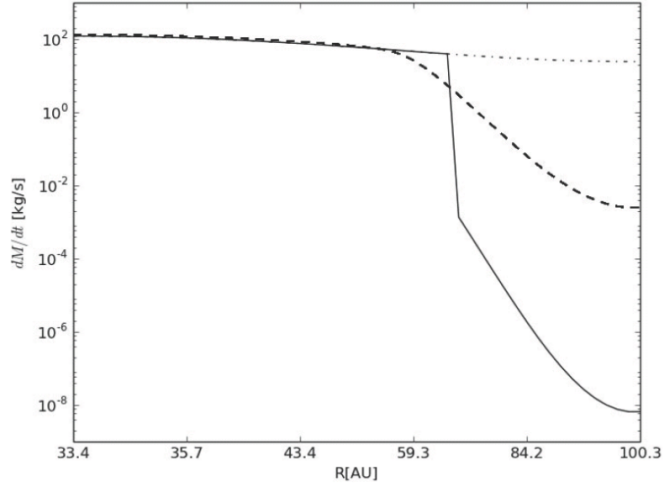

Fig. 4 2007OR10: estimated mass loss rate, *(dM/dt)*, for $N_2$ atmosphere containing a Pluto-like fraction of $CH_4$ vs. radial distance, *R*, from the sun in AU using $\rho$ = 1.8g/cm$^3$ as in Fig. 3; (solid line) using the larger of *(dM/dt)$_U$* from Eq. 3 and *(dM/dt)$_S$* with A = 0.67 using results in Fig. 2: the mass loss rate averaged over an orbit is *<dM/dt>* ~ *26kg/s*; (dot-dashed line): same using *A* = 0.1 indicating a huge change in *(dM/dt)$_S$* with *<dM/dt>* ~ *44kg/s*; (dashed line) *(dM/dt) = (dM/dt)$_S$ + f$_{UV}$ (dM/dt)$_U$* with *f$_{UV}$ =1- exp(-α N$_0$/N$_c$)* and *α* =10: difference in *<dM/dt>* from solid line is a few percent.

Integrating over each KBO's orbit, the orbital-averaged mass loss rate, *<dM/dt>*, is

$$<dM/dt> = \int (dM/dt)\, dt\, /\, \tau_{KBO} = \int (dM/dt)\, (R^2/2\pi\, ab)\, d\theta \qquad (4)$$

In Eq. 4, *a* and *b* are the semi-major and minor axes of the orbit, *θ* the angular position in the orbit (i.e., the true anomaly), $\tau_{KBO}$ the orbital period and *(dM/dt)* the mass loss rate at each *R*. Although Eris and 2007OR10 have similar orbital ranges, the effect of the surface albedo for the much more massive Eris is a few percent, but that is not the case for 2007OR10 as seen from Fig. 4. Although the effect of the gradual transition (dashed line) is only a few percent, the orbital-averaged $N_2$ loss rate, *<dM/dt>*, increases from 26kg/s for *A* = 0.67 (solid line) to 44kg/s for *A* = 0.1 (dot-dashed line).

In Table 2 we summarize orbital-averaged loss rates for all of the KBOs in Table 1. Estimates of escape driven *only* by surface heating are given in the SJ approximation used in SB, *<dM/dt>$_{SJ}$*, and then improved upon, *<dM/dt>$_S$*, using the analytic form in Fig. 2a & b. Results are also shown for loss driven by upper atmosphere heating, *<dM/dt>$_U$*, ignoring the thin-atmosphere cut-off. Orbital-averaged loss rates, *<dM/dt>*, are then given using the dominant process at each *R* and accounting for the thin atmosphere cut-off, $N_c$. For each KBO we use the mass, radius and orbital properties from Table 1. The results in Table 2 are for $N_2$ escape *prior* to loss of the primordial nitrogen inventory, so that we assume a relatively bright surface with *A* ~ 0.67 as seen in Table 1 for Pluto. For Makemake, 2007OR10, and Sedna we use both the upper and lower estimates of the density. Not surprisingly, the estimated loss rate due to surface heating is more sensitive to the KBO density than is loss due to upper atmosphere heating. Finally, for Charon orbiting Pluto, we reduce the gravitational binding to 0.91*U(r$_c$)* assuming escape occurs if the molecules reach its Hill sphere.

For a number of the KBOs the SJ approximation is seen in Table 2 to hugely overestimate the loss $N_2$ due to solar heating of the surface. With the exception of Sedna, which has an extremely thin atmosphere in its present orbit, it is seen that the upper atmospheric heating dominates the net mass loss for all KBOs studied even for a thick early $N_2$ atmosphere on Charon. As in SB we estimate the net loss of $N_2$ assuming that the trajectory and properties are unchanged for most of the lifetime of the KBO. If we assume each KBO is in its present orbit for $\tau \sim 4.5$Gyr ($1.4\times10^{17}$s) and the initial mass fraction of $N_2$ is $f_{N2}$, the fractional loss, $f_{loss}$, of the initial inventory of $N_2$ is

$$f_{loss} = [<dM/dt> \tau] / [f_{N2} M_{KBO}] \qquad (5)$$

Using Eq. 5 implies that all of the $N_2$ has access to the surface by diffusion from the interior, which is probably unlikely. Values for $f_{loss}$ are given in Table 2 using the mass fraction of nitrogen suggested in SB, $f_{N2} = 0.0074$. Unlike in SB the present estimates indicate that Charon is the only body that has clearly lost its full inventory with the approximate loss time given in brackets.

**Table 2 Orbital-Averaged Mass Loss Rates* [kg/s]**

| KBO | $<dM/dt>_{SJ}$ | $<dM/dt>_S$ | $<dM/dt>_U$ | $<dM/dt>$ | $f_{loss}$ | $\Delta r_{KBO}$(km) |
|---|---|---|---|---|---|---|
| Eris | $1.80\times10^{-28}$ | $4.6 \times 10^{-28}$ | 30. | 16. | 0.018 | 0.13 |
| Triton | $1.1\times10^{-24}$ | $9.7 \times 10^{-24}$ | 130. | 130. | 0.11 | 1.1 |
| Pluto | $1.4\times10^{-16}$ | $1.2 \times10^{-15}$ | 83. | 83. | 0.11 | 0.7 |
| Makemake | 12. - 0.013 | 4.1- 0.078 | 81. - 62. | 81. – 62 | 0.58 - 0.34 | 1.2 - 0.7 |
| Quaoar | $1.4\times10^3$ | 11. | 64. | 64. | 0.88 | 2.4 |
| 2007 OR10 | $5.2 - .043\times10^3$ | 3.1 - 1.4 | 49. - 38. | 29. - 22. | 0.31 - 0.18 | 0.77 - 0.55 |
| Charon | $1.8\times10^5$ | 17. | 110. | 110 | All {3.5Gyr} | 2.4 |
| Sedna | $8.4 - 0.1\times10^{-3}$ | $(16.-2.6)\times10^{-3}$ | 15. - 11. | $(16 - 2.6)\times10^{-3}$ | 0.018 | 0.13 |

* Orbital-averaged loss rates for a Pluto-like $N_2$ atmosphere containing a small fraction of $CH_4$, calculated using parameters from Table 1 but with $A = 0.67$: results for Charon are corrected for escape due to its Hill sphere and for Makemake, 2007 OR10, and Sedna are given for low and high densities; $<dM/dt>_{SJ}$: loss rate using the surface Jeans estimate in Eq. 1b as in SB; $<dM/dt>_S$: the $<dM/dt>_{SJ}$ corrected using the rough analytic estimate to the ratios in Fig. 2 a & b; $<dM/dt>_U$: loss rate induced by upper atmosphere heating in the EL approximation in Eq. 3 *not* requiring $N_0 > N_c$, for heating by $CH_4$ absorption; $<dM/dt>$: net orbital-averaged loss rate using the larger of $<dM/dt>_S$ and $<dM/dt>_U$ requiring that $N_0 > N_c$; $f_{loss}$: fraction lost of $N_2$ in Eq. 5 for $\tau = 4.5$Gyr and the initial mass fraction of $N_2$ from SB, $f_{N2}$ =0.0074; {} time to lose the primordial $N_2$ if it all eventually diffuses to the surface; $\Delta r_{KBO}$: rough radial change assuming compaction during loss of $N_2$, $\Delta r = \{\tau <dM/dt> / [4\pi \rho_{N2} R_{KBO}^2]\}$ with $\rho_{N2} = 1.0$ gm/cm$^3$.

Before discussing the relevance of the results in Table 2, we note that the higher EUV/UV fluxes in the early solar system can significantly decrease the time for a KBO to lose its primordial $N_2$. Ribas et al. (2005) showed that the EUV/UV radiation could be roughly scaled by time, *t*, as $F_0 (4.56\text{Gyr}/t)^y$ with $F_0$ the present flux. For Lyman-α they estimated $y \sim 0.72$ and for the radiation in the 92-110 nm, range $y \sim 0.85$. For the EUV in the 80-100 nm range absorbed by $N_2$, they estimated $y \sim 1.27$. If we assume the KBO's orbit, absorption efficiencies and heating efficiencies do not change significantly, as assumed in estimating $f_{loss}$, then the effective flux averaged from ~ 0.1Gyr to the present is enhanced by about a factor of 2.3 for the Lyman-alpha absorbed by $CH_4$ and a factor of ~ 6.7 for the EUV contribution to the energy absorbed by $N_2$. If we assume a KBO is in its orbit for nearly the lifetime of the solar system then the effective *Q* at 30AU averaged over that lifetime increases to $3.2\times10^{-3}$ ergs/cm$^2$/s and $1.9\times10^{-3}$ ergs/cm$^2$/s respectively for $CH_4$ and $N_2$. The resulting heat flux at Pluto would be ~ 3 times that typically used for solar medium at 30AU, enhancing the loss rate by a factor of ~3. In which case Pluto would have retained only ~60% of its initial $N_2$ and Makemake would likely lose all of its $N_2$. Of course, these bodies were likely *not* in their present orbits for the lifetime of the solar system nor would the albedo remain fixed during the evolution of the atmospheres, points that will be examined in the future.

To more clearly indicate the importance of the changes calculated here, we repeat the exercise in Fig. 1 in SB. They calculated the equivalent temperature of a body in a circular orbit that would have lost its primordial $N_2$ atmosphere in the lifetime of the solar system using a common average density, $\rho_{KBO} = 1.8 g/cm^3$. Assuming the SB estimate, $f_{N2} = 0.0074$, is correct, the heavy and light solid lines in Fig. 5 are the results based on their approximation in Eq. 1b and based on the upper atmospheric heating effect using Eq. 3 respectively. For the two models acting alone, bodies to the right of these lines would retain some of their $N_2$ and those to the left would have lost their $N_2$. Surprisingly it is seen that the diameters and effective temperatures for loss of the total inventory of $N_2$ is not hugely different for the two processes acting alone. But this is fortuitous as these lines depend critically on the parameters chosen. For instance, reducing $f_{N2}$ by a factor of ten, one obtains the dashed lines. It is seen the EL line changes significantly but the SJ does not. Therefore, even if the SJ model in SB was a reasonable lower bound, using it to categorize the KBOs is not useful, as the outcome is very insensitive to the initial conditions.

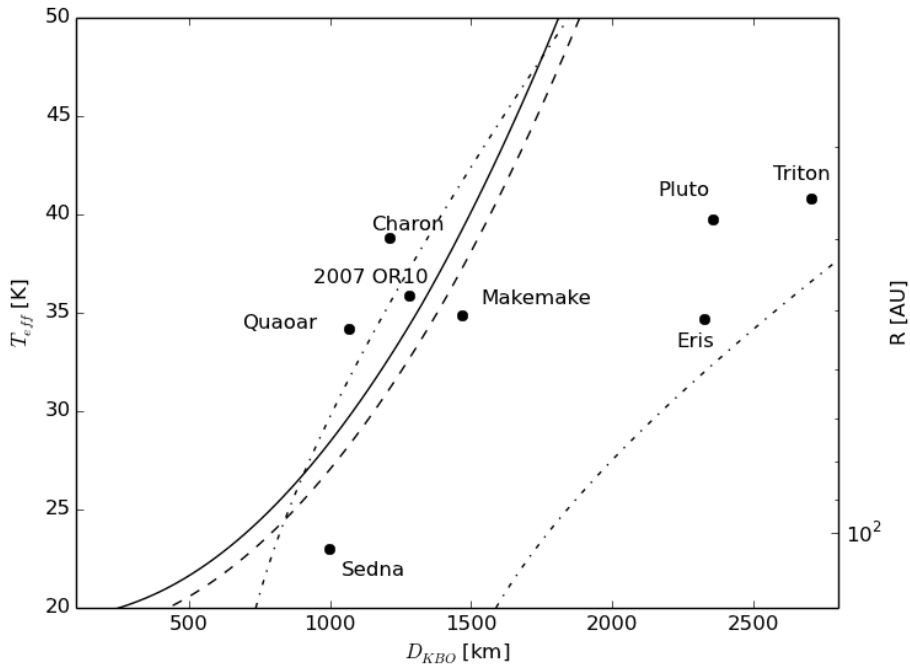

Fig 5. As in Fig. 1 in SB: lines are for KBOs of diameter $D_{KBO}$ (= $2r_0$) with equivalent temperature, $T_{eff}$, in a circular orbit to lose its primordial $N_2$ in 4.5 Gyrs. To the right of lines $N_2$ is retained, to the left it is lost. Using $f_{N2} = 0.0074$ as in SB and $f_{loss} = 1$ in Eq. 5: (solid line) SJ estimate from SB, (dash-dot on left) using the EL estimate in Eq. 3 with $A = 0.67$ to calculate $T_{eff}$. For $f_{loss} = 0.1$: (dashed) using SJ estimate in SB, (dash-dot) on right for EL estimate in Eq. 3.

**Summary**

We have presented new estimates of the loss of the primordial nitrogen from a number of large KBOs including the effect of both surface and atmospheric heating guided by recent simulations. The estimates in Table 2 are based on the assumption that the volatile fraction of $N_2$ assumed in SB is roughly correct, that all of the $N_2$ can eventually diffuse to the surface from which it can sublime into the atmosphere, that the presence of $CH_4$ in the early, predominantly $N_2$, atmospheres affects only the heating rate, that orbital parameters did not change for ~ 4.5 Gyr, that other loss processes, such as large impacts, are ignored, and that $A \sim 0.67$ during the period the KBO retains a significant $N_2$ atmosphere. Allowing for these rather drastic assumptions, the rates estimated in Table 2 indicate that earlier calculations were inaccurate for a number of large KBOs and that loss due to upper atmosphere heating is critical.

These calculations were carried out not only to improve upon previous estimates, but to gain insight into whether or not escape of $N_2$ can help understand observed KBO properties. The largest KBO's in Table 1 are seen to have much larger mass loss rates than those estimated earlier, losing as much as a kilometer of material, possibly affecting the morphology of their surface features. However, they still retain a significant fraction of their initial $N_2$ inventory, consistent with observations. Therefore, precipitation of a bright $N_2$ frost appears to correlate with their higher bond albedos, as noted in SB. Even if we account for the enhancement in the EUV/UV flux in the early solar system, Eris, Triton and Pluto would retain much of their initial $N_2$ inventory. On the other hand, if we include the enhancement in the UV in the early solar system, Makemake, which also has a relatively bright surface ($A \sim 0.62$), would likely have lost its full component of primordial $N_2$, contrary to what is suggested by observations.

The very low albedos ($A < 0.1$) on Quaoar and 2007 OR10 have been attributed to radiation-induced production of hydrocarbons that dominate the surface when the primordial $N_2$ is lost. Of course, as at Titan, the UV photolysis of $CH_4$, which heats the upper atmosphere, also produces hydrocarbons and a precipitate while the KBO still retains an $N_2$ atmosphere. By scaling to results for Titan (e.g., Strobel 1974; Atreya et al. 2006), the mass loss rate due to photolysis and precipitation of $CH_4$ products is smaller than $<dM/dt>_U$ in Table 2, but might affect the surface reflectance. However, as long as the $N_2$ component is robust, we presume that condensation covers much of this precipitate with a frost. After most of the $N_2$ eventually escapes, there might be a residual $CH_4$ atmosphere with a surface that becomes increasingly dark in the visible due to the precipitates or direct irradiation carbon species on the surface (e.g., Johnson 1989). Although this picture is likely correct, the complete loss of the primordial $N_2$ inventory is not easily explained by the results in Table 2. Allowing for a contribution from the enhanced UV in the early solar system, this picture *is* consistent with the results for Quaoar, although observations suggest the possible presence of some $N_2$. However, it is seen that 2007OR10, which has a very small albedo ($A \sim 0.06$), might not have lost its full $N_2$ inventory. That is, in the model described here complete loss would require the most extreme assumptions: the lowest density in Table 1, a low albedo as in Fig. 4, and being in its present orbit for nearly the lifetime of the solar system experiencing the enhanced UV/EUV flux. Although this is all possible, more detailed modeling than that described here is required for 2007OR10. Because the ratio $<dM/dt>_S / <dM/dt>_U$ based on Table 2 is largest for this intermediate-sized body, synergy between the surface and upper atmospheric heating could be important, an aspect we will be examining. However, unlike Quaoar, 2007OR10 is in a very eccentric orbit. During the period of time that 2007OR10 appears to retain most of its nitrogen component and a bright surface, the $N_2$ atmosphere would collapse at ~70AU. Beyond this point the surface becomes directly exposed to charged particle radiation, especially as it approaches the terminator shock at 80AU and the heliopause at ~100AU. This exposure could gradually reduce its albedo, enhancing the loss rate as indicated in Fig. 4. In this regard, it is also interesting to compare 2007OR10 to Eris, which has a similarly eccentric orbit and, in its present location (~100AU), has an albedo smaller than the other large objects. As they approach aphelion, they both cross the $N_0 > N_c$ solid line in Figs. 3 beyond which escape due to surface heating can dominate. But that escape rate is strongly affected by their large mass difference and, as seen in Fig. 3, any darkening increases the period of time which loss due to upper atmospheric heating rate dominates. Whereas the contributions in Table 2 to escape from Eris are small, this is not the case for 2007OR10 as seen in Fig. 4. Therefore, the evolution of the surface properties should be treated along with atmospheric escape, work that is now in progress.

Based on the albedos and loss rates, there are two outliers, Charon and Sedna. In its present orbit, Sedna's atmosphere is always thin and it spends much of its time outside the heliosphere. That is, although Sedna retains its primordial $N_2$, it is primarily as frozen $N_2$ on the surface, along with $CH_4$, with only a very tenuous atmosphere. Therefore, the surface reflectance is affected by the long-term irradiation of the mixed surface ice. The results in Table 2 also indicate that, if the primordial volatiles eventually all diffuse to the surface, Charon would have lost its initial $N_2$ inventory ~ 1Gyr ago, as is likely also the case for its $CH_4$ inventory. Charon's surface is, therefore, dominated by water ice, possibly having exposed regions with trapped nitrogen-containing molecules (Cook et al. 2007; Neveu et al. 2015). Such molecules could be from residual primordial nitrogen or nitrogen delivered by comets (Stern et al 2014). However, Pluto's escaping atmosphere continuously delivers $N_2$, which forms a thin atmosphere over the warmest regions but accumulates as a frozen layer in the cold regions until exposure to the sun (Tucker et al. 2014) a concept that will be tested during the New Horizon encounter.

We have shown here that the connection between the surface properties and atmospheric escape is likely much more subtle and interesting than suggested earlier. As knowledge of the physical properties of the KBOs, their orbital history and the initial volatile inventory improve the methods for estimating atmospheric loss given here can provide guidance when describing the volatile history. However, much more detailed simulations are needed on individual KBOs, taking both the surface and upper atmosphere heating into account, as well as the fate of $CH_4$. Now that it is clear that the upper atmospheric heating is critical, such simulations are in progress.

**Acknowledgements**

The work at Virginia was supported by grants from NASA's Outer Planet Research and Planetary Atmospheres Programs.